\documentclass[11pt]{article}
\usepackage{a4,amsmath,amsfonts,graphicx}
\usepackage{multirow}
\usepackage{natbib}
\usepackage{setspace}
\usepackage{ctable,rotating,colortbl,multirow}
\usepackage{lineno}
\usepackage{indentfirst}

\title{Efficient and Robust Estimation for a Class of Generalized Linear Longitudinal Mixed Models}
\author{Ren\'{e} Holst\\ rho@aqua.dtu.dk \\
National Institute of Aquatic Resources\\ Technical University
of Denmark\\Box 101, DK-9850 Hirtshals, Denmark \vspace{0.3cm}
\\
\vspace{0.3cm}
and \\
Bent J\o rgensen\\bentj@stat.sdu.dk\\
Department of Statistics\\University of Southern Denmark\\
Campusvej 55, DK-5230 Odense M, Denmark}
\date{}
\begin{document}
\newcommand{\tr}{\text{tr}}
\newcommand{\mCov}[2]{\text{cov}\left(#1,\,#2\right)}
\newcommand{\mVar}[1]{\text{var}\left(#1\right)}
\newcommand{\mE}[1]{E\left(#1\right)}
\newcommand{\Tw}[3]{\text{Tw}_{#1}\left( #2,\,#3\right)}
\newcommand{\aplus}[1]{\text{$\alpha_+^{#1}$}}
\newcommand{\AR}[1]{\textsc{ar}\ensuremath{\left(#1\right)}}
\newcommand{\MA}[1]{\textsc{ma}\ensuremath{\left(#1\right)}}
\newcommand{\gee}{\text{generalized estimating equations }}
\newcommand{\glmm}{\text{generalized linear mixed models}}
\newcommand{\cd}{\ensuremath{\negmedspace\cdot\negmedspace}}
\newcommand{\mbf}[1]{\ensuremath{\text{\boldmath${#1}$}}}
\renewcommand{\top}{\textrm{\scriptsize T}}
\newcommand{\emp}{\text{\scriptsize Emp}}
\label{firstpage}
\maketitle
\pagebreak
\noindent
{\bf Abstract}\\
We propose a versatile and computationally efficient estimating equation
method for a class of hierarchical multiplicative generalized linear mixed models
with additive dispersion components, based on explicit modelling of
the covariance structure. The class combines longitudinal and random effects
models and retains a marginal as well as a conditional interpretation.
The estimation procedure combines that of generalized estimating equations for the regression
with residual maximum likelihood estimation for the association parameters. This avoids the multidimensional
integral of the conventional generalized linear mixed models likelihood and allows an extension of the
robust empirical sandwich estimator for use with both association and
regression parameters. The method is applied to a set of otolith data,
used for age determination of fish.\\

\noindent
\textbf{Key words}:
Bias correction; Best linear unbiased predictor; Crowder weights; Dispersion
components; Generalized estimating equation; Nuisance parameter
insensitivity; Pearson estimating function; Residual maximum likelihood; Tweedie distribution.
\section{Introduction}
Ever since \cite{Nelder1972} introduced generalized linear models for independent data,
there has been a steady development of methods for analysis of non-normal correlated data.
This development was accelerated by the introduction of generalized estimating equations \citep{Liang1986}
and generalized linear mixed models \citep{Schall1991,Breslow1993,Wolfinger1993}.
These two types of models differ conceptually and computationally as reflected in the
conventional distinction between marginal and conditional models.

In practice, however, one is often faced with a combination of longitudinal and random
effects, where neither of the two, on their own, are adequate. In this light, it is somewhat of an enigma why generalized estimating equations and generalized linear mixed models have continued to evolve along separate paths. With few exceptions
\citep{McCulloch2001,Diggle2002,Fitzmaurice2004}, the literature is clearly divided into two separate strands.
\cite{Pinheiro2000} combined the two approaches for normal data.
\cite{Ziegler1998} and \cite{Hall2001} summarized the first decade of developments for generalized estimating equations
and recent contributions include for example \cite{Hardin2003},
\cite{Wang2004}, \cite{Coull2006}
and \cite{Wang2007a}.
For generalized linear mixed models, we refer to recent monographs, such as \cite{Verbeke2000}, \cite{Lee2006} and references therein.

In the present paper we propose a versatile and computationally efficient
method for generalized linear longitudinal mixed models. The estimating equations used for the regression parameters are similar in style to those known from conventional generalized estimating equations, whereas Pearson estimating equations are used for the association parameters.
The method combines longitudinal and random effects models
while retaining a marginal as well as a conditional interpretation.
A serial correlation structure is employed within clusters,
but unlike the state space models of \cite{Jorgensen1999} and \cite{Jorgensen2007},
where the latent process is non-stationary, our approach is based on
a stationary latent process defined by means of a linear filter.

In order to avoid the intricate efficiency considerations associated with conventional generalized estimating equations,
see eg. \cite{Wang2003}, we emphasize explicit modelling of the covariance structure
based on second-moment assumptions for the data-generating process. An example of the utility of this approach is the twin data analysis by \cite{Iachina2002}, where the correlation
between the two twins in a pair was estimated on the basis of a generalized estimating equations model.
Nevertheless, the method allows the use of a working correlation structure
and we extend the robust empirical sandwich estimator
to handle the asymptotic variance for both regression and association parameters.

Our estimation of the association parameters compares to that of residual maximum likelihood
by means of the bias-corrected Pearson estimating equations of \cite{Jorgensen2004},
following earlier work by \cite{McCullagh1990} and \cite{Hall1998}.

The estimating equations are solved by an efficient Newton scoring algorithm, thereby
circumventing the problems associated with the multidimensional integral
defining the likelihood in conventional generalized linear mixed models approaches such as \cite{Schall1991},
\cite{Breslow1993} and \cite{Wolfinger1993}.

The models covered by our approach are defined hierarchically via conditional Tweedie distributions,
much like the multiplicative mixed effects models of \cite{Ma2007} and \cite{Ma2009}.
This leads to a multiplicative mean structure with a corresponding additive
decomposition of the variance into dispersion components, thereby
retaining much of the simplicity of classical linear models.
The Tweedie family provides a flexible class of models for both positive continuous data,
count data and positive continuous data with a point mass at zero \cite[Ch.~4]{Jorgensen1997},
see also \cite{Jorgensen2007}.
\section{Model}
\label{s:model}
\subsection{Model specification}
We now introduce the main type of our approach followed by a discussion  of their covariance structure, which is crucial for the interpretation and estimation of the models. The model is based on the class of Tweedie exponential dispersion models \citep[Ch.~4]{Jorgensen1997}. A Tweedie variable $Y \sim  \Tw{r}{\mu}{\sigma^2}$ is characterized by having $\mE{Y}=\mu$ and $\mVar{Y}=\sigma^2\mu^r$. Special cases include the normal ($r=0$), Poisson ($r=1$ and $\sigma^2=1$) and gamma ($r=2$) families. The case $1<r<2$ correspond to compound Poisson distributions, which are non-negative and continuous, except for a positive probability at zero.

For a given cluster $i$ and time $t$ consider a response vector
$Y_{it}$  with conditionally independent Tweedie distributions.
For ease of presentation we
consider a balanced design with $T$ equidistant observation times
common to all $I$ clusters. The model is readily adapted to ragged structures.

The cluster random effect is represented by independent latent Tweedie variables,
\begin{equation}
 Z_i \sim \Tw{r_1}{1}{\sigma^2} \label{eq:Zi},
\end{equation}
where $r_1 \ge 2 $ in order to make $Z_i$ positive.
Given the cluster variable $Z_{i}$, we consider a latent process based on Tweedie noise,
\begin{equation}
Z_{it}\mid Z_i=z_i \sim \Tw{r_{2}}{\aplus{-1}z_i\,}{\aplus{r_{2}}\omega^2z_i^{1-r_{2}}},\label{eq:Zitj}
\end{equation}
where the $Z_{it}$ are assumed conditionally independent given the
variables $Z_{i}$, and again $r_{2} \ge 2 $. Here $\alpha_+=\sum_{k=0}^{\infty}\alpha_k < \infty$ with
$\alpha_0=1$ and $\alpha_k \in [0,1)$ for $k>0$ .
The coefficients $\alpha_k$ determine a conditionally weakly stationary latent process $Q_{it}$,
defined by the linear filter
\begin{equation}
Q_{it}=\sum_{s=0}^\infty \alpha_s Z_{i\,t-s}. \label{eq:Qitj}
\end{equation}
By way of construction of the middle layer of the latent process $Q_{it}$ has mean $1$.
At the observation level we assume
\begin{equation}
Y_{itj}\mid \mbf Q_{i*}=\mbf q_{i*}\sim \Tw{r_{3}}{\mu_{it}
q_{it}}{\rho^2 q_{it}^{1-r_{3}}}, \label{eq:Yitj}
\end{equation}
with conditional independence given $Q_{i*}$. Here the $*$
notation denotes the vector obtained by letting the corresponding index run,
so that $\mbf Y_{i*}=(Y_{i1}, \ldots, Y_{iT})^{\top}$ and so on. By definition of the Tweedie variance function we obtain linearity in the conditioning variable of the mean and variance, so that $\mE{Y_{it}\mid \mbf Q_{i*}}=q_{it} \mu_{it}$ and $\mVar{Y_{it}\mid \mbf Q_{i*}}=q_{it}\rho^2 \mu_{it}^{r_{3}}$. This property is essential for the derivation of the covariance structure and the estimating functions below. In the Poisson case ($r_{3}=1$) it is convenient to let the dispersion parameter $\rho^2>0$ accommodate
potential over- or under-dispersion.

A variant of the model, applicable to normal response variables, assumes normal zero mean random cluster effects, $Z_i \sim N(0,\sigma^2)$, replacing (\ref{eq:Zi}). The noise process (\ref{eq:Zitj}) is then assumed to be Gaussian, $Z_{it}\mid Z_i=z_i \sim N(z_i,\omega^2)$, while maintaining the filter (\ref{eq:Qitj}) as above.
At the response level (\ref{eq:Yitj}) is replaced by an additive structure with identity link function,  $Y_{it}\mid \mbf Q_{i*}=\mbf q_{i*}\sim N(\mu_{it}+q_{it},\rho^2)$. Since the structure is linear, the corresponding covariance structure is easily derived.

The marginal means may depend on covariates $\mu_{it}=\mu_{it}\left(\mbf  x_{it};\mbf \beta\right)$, where $\mbf \beta$ denotes a vector of regression parameters. With positive data, the log link is a suitable choice, providing a natural interpretation of the regression parameters. Furthermore the log link, along with the multiplicative structure, enables easy comparison with conventional generalized linear mixed models, where the random effects enter as a term in the linear predictor: $\eta_{it}=\log\left(\mu_{it} q_{it}\right)=
\mbf  x_{it}^{\top}\mbf \beta+\log\left(q_{it} \right)$.
\subsection{Covariance structure}
The marginal variance-covariance matrix of the observation vector $\mbf Y_{**}$ is crucial for the inference and estimation procedures. Detailed derivations are found in Appendix 1.

The covariance between two given observations within the $i$th cluster is
\begin{equation*}\mCov{Y_{it}}{Y_{it'}}=
\sigma^2\mu_{it}\mu_{it'}+\omega^2\mu_{it}\mu_{it'}
\sum_{s=0}^\infty \alpha_s \alpha_{s+|t-t'|}
\,+\,\delta_{t}^{t'}\!\rho^2\mu_{it}^{r_{3}}
\label{eq:CovYitjYit'j'},
\end{equation*}
where $\delta_i^{i'}$ is the Kronecker delta, being $1$ for $i=i'$ and zero otherwise. It is an important property of the model that the covariance 
does not depend on $r_1$ and $r_2$. From this we now derive a matrix expression for $\mVar{\mbf Y_{**}}$.

First we consider the latent process correlation matrix, $K\left(\alpha\right)$, with $tt'$th entry $\sum_{s=0}^\infty \alpha_s \alpha_{s+|t-t'|}$. Next let $\mbf 1_T$ denote the $T$-vector of $1$s.
In matrix notation the variance-covariance matrix of the response vector for the $i$th cluster
may then be expressed as
\begin{align}
\label{eq:var.Y.i.mat}
\mVar{\mbf Y_{i*}}&=\mbf \mu_{i*} \mbf \mu_{i*}^{\top} \odot
\left\{\sigma^2\
\mbf 1_T \mbf 1_T^{\top}+ \omega^2\,\mbf K\left(\mbf \alpha\right)\right\}+\rho^2\ \text{diag}\left(\mbf \mu_{i*}^{r_{3}}\right)     \notag \\
&=\sigma^2\,\mbf \mu_{i*}\mbf \mu_{i*}^{\top}
+\omega^2\,\text{diag}\left(\mbf \mu_{i*}\right)\mbf K\left(\mbf \alpha\right)\,\text{diag}\left(\mbf \mu_{i*}\right)
+ \rho^2\ \text{diag}\left(\mbf \mu_{i*}^{r_{3}}\right)
,
\end{align}
say, where $\odot$ is the Hadamard (elementwise) product \citep[~p. 45]{Magnus1999}.

Similar to conventional linear mixed models, the variance is decomposed into components of dispersion
corresponding to the different sources of variation. The covariance terms in (\ref{eq:var.Y.i.mat}) reflect the observation
error, the covariance within cluster and the variation  between clusters.

The models, accommodated by our approach, hence extend the range of possible serial correlation patterns compared with the conventional generalized estimating equations correlation structures usually considered. Particular covariance structures may be obtained by imposing restrictions on the linear filter parameter vector $\mbf \alpha$ or the dispersion parameters. Table \ref{tbl:CorrStructs} lists the more common covariance structures and the corresponding parameter restrictions.
\begin{table}[hptb]
\def~{\hphantom{0}}
\caption{Standard covariance structures. The \MA{p}-type and \AR{p}-type refer to the latent process correlation structure conditionally on the cluster random effects.}\label{tbl:CorrStructs}
\begin{center}
\begin{tabular}{c p{9cm}}
\toprule
Covariance structure & Parameter restrictions \\ \hline
Independent & $\omega_j^2=\sigma^2=0$.\\
Exchangeable & $\omega_j^2=\rho^2_j=0 \text{ and } \alpha_s=0 \text{ for }\, s>0$. \\
\MA{p}-type & $\alpha_s=0$ for $s>p$.\\
\AR{p}-type & For $p=1\ \ \alpha_s=\alpha^s$. For $p>1$ the $\alpha_s$ are given by the Yule-Walker equations.\\
GLMM & $\omega_j^2=0 \text{ and } \alpha_s=0 \text{ for }\, s>0$.\\
\bottomrule
\end{tabular}
\end{center}
\end{table}
\section{Estimation}
\label{s:estimation}
\subsection{General issues}
The set of parameters $\mbf \theta$ to be estimated is naturally partitioned into regression
and association parameters, $\mbf \theta=\left(\mbf \beta^{\top}, \mbf \gamma^{\top}\right)^{\top}$, where the regression parameters
$\mbf \beta$ usually are those of interest whereas the association parameters $\mbf \gamma$, containing dispersion
and correlation parameters, are often considered nuisance parameters. For estimation of the parameters we use a set of corresponding estimating functions denoted
$\mbf \psi=\left(\mbf \psi^{\top}_{\mbf \beta},\mbf \psi^{\top}_{\mbf \gamma}\right)^{\top}$.

The estimating function for the regression parameters is
\begin{align}
\mbf \psi_{\mbf \beta}&=\sum_{i=1}^{I}\mbf D_i^{\top}
\mbf C^{-1}_i \left(\mbf Y_i - \mE{{\mbf Y_i}} \right),
\label{eq:psi.xi.blup}
\end{align}
where $\mbf D_i=\nabla_{\mbf \beta}\mE{\mbf Y_i}=\partial \mE{\mbf Y_i}/\partial \mbf \beta^{\top}$ and $\mbf C_i=\mVar{\mbf Y_i }$. Although  (\ref{eq:psi.xi.blup}) is similar to the well known estimating function for the regression parameters from the conventional generalized estimating equations framework \citep{Liang1986}, it corresponds to using the model covariance matrix (\ref{eq:var.Y.i.mat}) rather than the working covariance matrix. The conventional generalized estimating equations working covariance matrix is built around the working correlation matrix $\mbf R(\mbf \alpha)$ so that
\begin{equation}\mVar{\mbf Y_i}=\phi \mbf A_i^{1/2}\left(\mbf \mu_i\right)\mbf R\left(\alpha\right)\mbf A_i^{1/2}\left( \mbf \mu_i\right)\label{eq:working.covar},\end{equation}
where $\phi$ is a dispersion parameter,
$\mbf A_i\left(\mbf \mu_i\right)=\text{diag}\left\{v\left(\mbf \mu_{i*}\right)\right\}$ and $v\left(\cdot\right)$ is the variance function. In contrast, we emphasize the decomposition of the variance into components of dispersion and associate the process correlation matrix $\mbf K(\mbf \alpha)$ with an appropriate level in the hierarchy.

We use Pearson estimating functions for the estimation of the association parameters $\mbf \gamma=\left(\gamma_1,  \ldots, \gamma_N\right)^{\top}$.  The entire vector of functions is denoted $ \mbf \psi_{\mbf \gamma}=\left(\psi_{\gamma_1}, \ldots, \psi_{\gamma_N}\right)^{\top}$, where $N=3+M$ and $M=\text{dim}\left(\mbf \alpha\right)$ and with the $n$th component given by
\begin{align*}
\psi_{\gamma_n}\left(\mbf \beta, \mbf \gamma\right)
&=\sum_{i=1}^I  \text{tr} \left\{\mbf  W_{in}\left(\mbf r_i \mbf r_i^{\top} -\mbf C_i \right)\right\}, \label{eq:psi.gamma.j}
\end{align*}
where $\mbf r_i=\mbf Y_i-\mE{\mbf Y_i}$ and $\mbf W_{in}$ are suitable weights. This form emphasizes the model covariance matrix in
contrast to the more conventional expressions of the Pearson estimating function \citep{Jorgensen2004}.

The estimating functions $\mbf \psi_{\mbf \beta}$ and $\mbf \psi_{\mbf \gamma}$ are explained in more detail below, in Section \ref{s:regparameters} and \ref{s:assocparams} respectively.
\subsection{Sensitivity}
\label{s:sensitivity}
\cite{Cox1987} studied parameter orthogonality in the likelihood framework corresponding to block diagonality of the Fisher information matrix. \cite{Jorgensen2004} studied the corresponding property of nuisance parameter insensitivity in an estimating equation context, by means of the sensitivity matrix, defined by
$\mbf S_{\mbf \theta}=E\left\{\nabla_{\mbf \theta}\mbf \psi(\mbf \theta)\right\}$, where $\nabla$ is the gradient operator.
The sensitivity matrix may be partitioned into blocks corresponding to $\left(\mbf \psi_{\mbf \beta}^{\top},\mbf \psi_{\mbf \gamma}^{\top}\right)^{\top}$ and $\left({\mbf \beta}^{\top},{\mbf \gamma}^{\top}\right)$ as follows:
\[\mbf S_{\mbf \theta}=\begin{bmatrix} \mbf S_{\mbf \beta}\left(\mbf \theta\right) &  \mbf S_{\mbf \beta \mbf \gamma}\left(\mbf \theta\right)\\
\mbf S_{\mbf \gamma \mbf \beta }\left(\mbf \theta\right)& \mbf S_{\mbf \gamma}\left(\mbf \theta\right)
\end{bmatrix}
=\begin{bmatrix} E\left\{\nabla_{\mbf \beta}\mbf \psi_{\mbf \beta}(
\mbf \beta, \mbf \gamma)\right\} &  E\left\{\nabla_{
\mbf \gamma}\mbf \psi_{\mbf \beta}(\mbf \beta, \mbf \gamma)\right\}\\
E\left\{\nabla_{\mbf \beta}\mbf \psi_{\mbf \gamma}(\mbf \beta,
\mbf \gamma)\right\} & E\left\{\nabla_{\mbf \gamma}\mbf \psi_{\mbf \gamma}(
\mbf \beta, \mbf \gamma)\right\}\end{bmatrix}.\]
Nuisance parameter insensitivity (for short denoted $\mbf \gamma$-insensitivity) is defined by the upper right-hand block $S_{\mbf \beta\mbf \gamma}\left(\mbf \theta\right)$ being zero. First of all this implies efficiency stable estimation of $\mbf \beta$, meaning that the estimation of $\mbf \gamma$ does not affect the asymptotic variance of $\widehat{\mbf \beta}$; see Section \ref{s:godambe}. Second, it simplifies the Newton scoring algorithm \citep{Jorgensen2004} as detailed below. Third, it implies that $\widehat{\mbf \beta}_{\mbf \gamma}$ varies only slowly with $\mbf \gamma$, where $\widehat{\mbf \beta}_{\mbf \gamma}$ is the estimate of $\mbf \beta$ for with $\mbf \gamma$ known. While nuisance parameter-insensitivity does not ensure asymptotic independence of $\widehat{\mbf \beta}$ and $\widehat{\mbf \gamma}$, it does ease the computation of the asymptotic variance of $\widehat{\mbf \beta}$.

Following \cite{Jorgensen2004} it is easily seen that $\mbf \psi_{\mbf \beta}$ is  $\mbf \gamma$-insensitive. In fact, from (\ref{eq:psi.xi.blup}) we see that
$\mbf \psi_{\mbf \beta}$ depends on $\mbf \gamma$ only via $\mbf C_i^{-1}$ and hence $\nabla_{\mbf \gamma}\mbf \psi_{\mbf \beta}\left(\mbf \beta, \mbf \gamma\right)$ has zero mean, i.e. $\mbf S_{\mbf \beta\mbf \gamma}\left(\mbf \theta\right)=0$.

In the rest of the paper we write  $\mbf S_{\mbf \beta}$ for $\mbf S_{\mbf \beta}\left(\mbf \theta\right)$ etc, whenever the meaning is unambiguous.
The remaining blocks of $\mbf S_{\mbf \theta}$ are detailed along with the estimating functions in Sections \ref{s:regparameters} and \ref{s:assocparams}
\subsection{Algorithm}
Calculation of the parameter estimates is achieved by the Newton scoring algorithm \citep{Jorgensen1996a} in which we update
the previous value of $\mbf \theta$ by
\[\mbf \theta^*=\mbf \theta-\mbf S^{-1}_{\mbf \theta}\mbf \psi\left(\mbf \theta\right).\]
By the regularity of $\mbf \psi$, the $\mbf \gamma$-insensitivity of $\mbf \psi_{\mbf \beta}$, and using simple matrix manipulations we may express the inverse of $\mbf S_{\mbf \theta}$ in blocks as follows:
\begin{equation}\mbf S_{\mbf \theta}^{-1}=\begin{bmatrix} \mbf S_{\mbf \beta}^{-1} &  \mbf 0\\
-\mbf S_{\mbf \gamma}^{-1}\mbf S_{\mbf \gamma \mbf \beta }\mbf S_{\mbf \beta}^{-1}& \mbf S_{\mbf \gamma}^{-1}
\end{bmatrix}. \label{eq:S.theta}
\end{equation}
The algorithm therefore splits into a $\mbf \beta$ step
\begin{equation}
\mbf \beta^*=\mbf \beta-\mbf S^{-1}_{\mbf \beta}\mbf \psi_{\mbf \beta}\left(\mbf \theta\right) \label{eq:newton.xi}
\end{equation}
and a $\mbf \gamma$ step
\begin{align}
\mbf \gamma^*&=\mbf \gamma+\mbf S_{\mbf \gamma}^{-1} \mbf S_{\mbf \gamma\mbf \beta }\mbf S_{\mbf \beta}^{-1}\mbf \psi_{\mbf \beta}\left(\mbf \theta\right)-\mbf S^{-1}_{\mbf \gamma}\mbf \psi_{
\mbf \gamma}\left(\mbf \theta\right) \notag\\
&=\mbf \gamma-\mbf S_{\mbf \gamma}^{-1}\left\{\mbf \psi_{\mbf \gamma}\left(\mbf \theta\right)-\mbf S_{\mbf \gamma\mbf \beta }
\mbf S_{\mbf \beta}^{-1}\mbf \psi_{\mbf \beta}\left(\mbf \theta\right)\right\} \label{eq:newton.gamma}.
\end{align}
Following \cite{Jorgensen2004} we insert $\mbf \beta^*$ from (\ref{eq:newton.xi}) into equation
(\ref{eq:newton.gamma}). Since equation (\ref{eq:newton.xi}) can be
rewritten as $-\mbf S_{\mbf \beta}^{-1}\left(\mbf \theta\right)\mbf \psi_{\mbf \beta}\left(\mbf \theta\right)=\mbf \beta^*-\mbf \beta$, this makes $\mbf S_{\mbf \beta}^{-1}\left(\mbf \theta^*\right)\mbf \psi_{\mbf \beta}\left(\mbf \theta^*\right)=0$,
where $\mbf \theta^*$ indicates $\mbf \beta^*$ is being used.
Consequently the modified $\mbf \gamma$ step becomes
\begin{equation}
\mbf \gamma^*=\mbf \gamma-\mbf S_{\mbf \gamma}^{-1}\mbf \psi_{\mbf \gamma}\left(\mbf \theta^*\right).\label{eq:newton.gamma2}
\end{equation}
Analogously we use the most recent estimate of $\mbf \gamma$ when
updating $\mbf \beta$ in (\ref{eq:newton.xi}). This is however of less importance, due to the slow variation
of $\widehat{\mbf \beta}_{\mbf \gamma}$ with $\mbf \gamma$. \cite{Jorgensen2004} coined the scheme of alternating
between (\ref{eq:newton.xi}) and (\ref{eq:newton.gamma2}) the \textit{chaser} algorithm, with reference to the asymmetric interdependence between $\mbf \beta^*$ and $\mbf \gamma^*$.
Our experience with the algorithm  confirms this property.
\subsection{Regression parameters $\mbf \beta$}
\label{s:regparameters} Following \cite{Ma1999}, \cite{Ma2003}
and \cite{Ma2007} we use best linear unbiased predictor for predicting the random effects.
The best linear unbiased predictor of a random variable $\mbf Q$ given the observed data
$\mbf Y$ is defined by \citep{Henderson1975,Ma1999}
\begin{equation}
\widehat{\mbf Q}
=\mE{\mbf Q}+\mCov{\mbf Q}{\mbf Y}\mVar{\mbf Y}^{-1}\left\{\mbf Y-\mE{\mbf Y}\right\}.\label{eq:blup.definition}
\end{equation}
The model specification using Tweedie distributions allows for
derivation of the joint score function $u \left(\mbf \theta;\mbf Y,\mbf Q\right)$ \citep{Ma2007}. We define unbiased
estimating functions $\mbf \psi_{\mbf \beta}$  by
substituting the random effects by their respective best linear unbiased predictors, i.e. \[\mbf \psi_{\mbf \beta}\left(\mbf \theta; \mbf Y\right)=u\left(\mbf \theta;
\mbf Y,\widehat{\mbf Q}\right).\]
It follows from (\ref{eq:blup.definition}) and the linearity of
$\mE{\cdot}$ and $\mCov{\cdot}{\mbf Y}$ that the best linear unbiased predictor of
$\mbf A \mbf Q+ \mbf B \mbf Y$ given $\mbf Y$ is $\mbf A \widehat{\mbf Q}+ \mbf B \mbf Y$, where $\mbf A$
and $\mbf B$ are matrices of suitable dimensions. Since $u$ is linear in
both the observed and the latent variables, $\mbf Y$ and $\mbf Q$,
we find that $\mbf \psi_{\mbf \beta}$ is the best linear unbiased predictor of the score function $u$ given the data. By (\ref{eq:blup.definition}) we therefore arrive at the conventional generalized estimating equations form expression (\ref{eq:psi.xi.blup})
\begin{equation}\mbf \psi_{\mbf \beta}=\mE{u}+\mCov{u}{\mbf Y} \mbf C^{-1}\left\{\mbf Y -\mE{\mbf Y}\right\}=\sum_{i=1}^{I}\mbf D_i^{\top}\mbf C^{-1}_i \left\{\mbf Y_i-\mE{\mbf Y_i} \right\}.\label{eq:psi.beta.as.blup}
\end{equation}
Here we have used the independence between clusters, along with the following Bartlett-type identity
\[\mbf D=\nabla_{\mbf \beta}\mE{\mbf Y}=\mE{\mbf Y\cdot u}=\mCov{\mbf Y}{u}.\]
From (\ref{eq:psi.xi.blup}) we furthermore obtain the sensitivity $\mbf S_{\mbf \beta}=\mE{\nabla_{\mbf \beta}\mbf \psi_{\mbf \beta}}$ and variability $\mbf V_{\mbf \beta}=\mVar{\mbf \psi_{\mbf \beta}}$ as
\begin{equation}
\mbf S_{\mbf \beta} =-\mbf D^{\top}\mbf C^{-1}\mbf D,  \quad \mbf V_{\mbf \beta}=\mbf D^{\top}\mbf C^{-1}\mbf D \label{eq:S.beta.and.V.beta}.
\end{equation}
The identity $\mbf V_{\mbf \beta}=-\mbf S_{\mbf \beta}$ is characteristic
for quasi-score functions. We therefore conclude that $\mbf \psi_{\mbf \beta}$ is optimal within the class of linear estimating functions. This also follows from (\ref{eq:psi.beta.as.blup}) as the best linear unbiased predictor is optimal among all linear predictors.
\subsection{Association parameters $\mbf \gamma$}
\label{s:assocparams}
Our approach is akin to that of \cite{Ma2007}, but deviates by allowing for correlation structures within clusters. Furthermore Ma and J\o rgensen used a closed form ad-hoc estimator for the association parameters
Our estimation of $\mbf \gamma$ is based on Pearson estimating functions, following the path of \cite{Hall1998} and \cite{Jorgensen2004}.
For $\gamma_n$ it is defined by
\begin{align}
\psi_{\gamma_n}\left(\mbf \beta, \mbf \gamma\right) &=\sum_{i=1}^I \left\{\mbf r_i^{\top} \mbf W_{in}\mbf r_i -\text{tr} \left(\mbf W_{in}\mbf C_i \right)\right\} \notag\\
&=\sum_{i=1}^I \left\{ \text{tr} \left(\mbf W_{in}\mbf r_i \mbf r_i^{\top} \right) -\text{tr} \left(\mbf W_{in}\mbf C_i \right) \right\} \notag\\
&=\sum_{i=1}^I  \text{tr} \left\{\mbf W_{in}\left(\mbf r_i \mbf r_i^{\top} -\mbf C_i \right)\right\}, \label{eq:psi.gamma.j}
\end{align}
where $\mbf r_i=\mbf Y_i-\mE{\mbf Y_i}$ and $\mbf W_{in}$ are
suitable weights. By linearity of $\mE{\cdot}$ and
$\text{tr}(\cdot)$ these estimating functions are clearly unbiased
since $\mE{\mbf r_i \mbf r_i^{\top}}=\mbf C_i$.

In the conventional generalized estimating equations framework the Pearson estimating function
hinges the estimation of association parameters on a working
correlation matrix used for defining $\mVar{\mbf Y_i}$ as shown in (\ref{eq:working.covar}).
In contrast, we emphasize the decomposition of the variance into components
of dispersion and associates the process correlation matrix
$\mbf K\left(\mbf \alpha\right)$ with an appropriate level in the
hierarchy.

For $\mbf W_{in}$ we use the weights proposed by \cite{Hall1998},
\begin{align*}
\mbf W_{in}&=-\frac{\partial \mbf C_i^{-1}}{\partial \gamma_n}=\mbf C_i^{-1}\frac{\partial \mbf C_i}{\partial \gamma_n}\mbf C_i^{-1}.
\end{align*}
In the normal case these weights lead to quasi-score  functions and in general they provide estimating functions
that resemble the structure of the estimating functions (\ref{eq:psi.xi.blup}) for
the regression parameters.

From (\ref{eq:psi.gamma.j}) we may derive the $\theta_m$-sensitivity of $\psi_{\gamma_n}$, namely
\begin{align*}
\mE{\frac{\partial}{\partial\theta_m}\psi_{\gamma_n}} &=E\left[\frac{\partial}{\partial\theta_m}\sum_{i=1}^I  \text{tr} \left\{\mbf W_{in}\left(\mbf r_i \mbf r_i^{\top}-\mbf C_i \right)\right\}\right] \notag\\
&=\sum_{i=1}^I  \text{tr} \left[\mbf W_{in}E\left\{\frac{\partial}{\partial\theta_m}\left(\mbf r_i \mbf r_i^{\top}-\mbf C_i \right)\right\}\right]\notag\\
&=-\sum_{i=1}^I  \text{tr} \left(\mbf W_{in}\frac{\partial \mbf C_i}{\partial\theta_m}\right).
\end{align*}
Here we have used that the derivatives of $\mbf r_i$ do not depend on data so
$E\left\{\left(\partial \mbf r_i/\partial \theta_m\right)\mbf r_i^{\top}\right\}=E\left\{\mbf r_i\left(\partial \mbf r_i^{\top}/\partial \theta_m\right)\right\}=0$.

The symmetry between the blocks $\mbf S_\gamma$ and $\mbf S_{\mbf \gamma\mbf \beta}$ is highlighted by the forms of the $nm$th entries
\begin{align}\left\{\mbf S_\mbf \gamma\right\}_{nm}&=-\sum_{i=1}^I  \text{tr}\left(\mbf C_i^{-1}\frac{\partial \mbf C_i}{\partial \gamma_n}\mbf C_i^{-1}\frac{\partial \mbf C_i}{\partial \gamma_m}\right) \label{eq:S.gamma}\\
\intertext{and}
\left\{\mbf S_{\mbf \gamma\mbf \beta}\right\}_{nm}&=-\sum_{i=1}^I  \text{tr}\left(\mbf C_i^{-1}\frac{\partial \mbf C_i}{\partial \gamma_n}\mbf C_i^{-1}\frac{\partial \mbf C_i}{\partial \beta_m}\right) \label{eq:S.gamma.beta}
\end{align}
respectively.
\subsection{Bias correction}
\label{s:biascorrection}
The estimation of nuisance parameters may be subject to bias \citep{McCullagh1990, Jorgensen2004}, caused by not taking
into account the degrees of freedom spent on estimating the regression parameters.

In the spirit of \cite{Godambe1960}, \cite{Heyde1997} and \cite{Jorgensen2004} we adjust the estimating function for bias rather than the estimate.
The corrected estimating function for $\gamma_n$ becomes
\begin{align*}
\breve{\psi}_{\gamma_n}\left(\mbf \beta,\mbf \gamma\right) & =
\psi_{\gamma_n}\left(\mbf \beta,
\mbf \gamma\right)+b_{\gamma_n}\left(\mbf \beta,\mbf \gamma\right)\\
&=\sum_{i=1}^I  \text{tr} \left\{\mbf W_{in}\left(\mbf r_i \mbf r_i^{\top}-\mbf C_i \right)\right\}
+\text{tr} \left\{ \left(\sum_{i=1}^I \mbf D_i^{\top} \mbf W_{in} \mbf D_i\ \right)\left(\sum_{i=1}^I \mbf D_i^{\top}
\mbf C_{i}^{-1} \mbf D_i\ \right)^{-1}\right\} \\
&=\sum_{i=1}^I  \text{tr} \left\{\mbf W_{in}\left(\mbf r_i \mbf r_i^{\top} -\mbf C_i \right)\right\}-\text{tr} \left(\bf J_{\mbf \beta}^{\left(\gamma_n\right)}\bf J_{\mbf \beta}^{-1}\right)
,
\end{align*}
where $\bf J_{\mbf \beta}^{(\gamma_n)}=\partial \bf J_{\mbf \beta}/\partial\gamma_n$.
The Godambe information $\bf J_{\mbf \beta}$, see Section \ref{s:godambe}, plays a role in the estimating equation context analogous to that of the Fisher information in the likelihood framework, with $\bf J_{\mbf \beta}^{-1}$ being the asymptotic variance of ${\widehat {\mbf \beta}}$. The penalty term $b_{\gamma_n}\left(\mbf \beta,\mbf \gamma\right)$ therefore represents the $\mbf \gamma$-dependency of $\bf J_{\mbf \beta}$, weighted by the precision of the estimate ${\widehat {\mbf \beta}}$. In this way it corrects for the effect upon $\psi_{\gamma_n}\left({\widehat{\mbf  \beta}}_{\mbf \gamma},\mbf \gamma\right)$ of using ${\widehat {\mbf \beta}_{\mbf \gamma}}$.

We note that $b_{\gamma_n}\left(\mbf \beta, \mbf \gamma\right)=\partial\log\left(\left|\bf J_\mbf \beta^{-1}\right|\right)/\partial \gamma_n$, which may be a more convenient form in some applications.

Since $b_{\gamma_n}\left(\mbf \beta,\mbf \gamma\right)$ does not depend on the data, we obtain the
$\mbf \gamma$- and $\mbf \beta$-sensitivity of $\breve{\mbf \psi}_{\mbf \gamma}\left(\mbf \beta, \mbf \gamma\right)$ by amending $\mbf S_{\mbf \gamma}$ and $\mbf S_{\mbf \gamma\mbf \beta}$ respectively, with the $\mbf \gamma$- and $\mbf \beta$-derivatives, of the penalty term, respectively. For the $nm$th entries we obtain
\begin{align}
\frac{\partial}{\partial \gamma_m}b_{\gamma_n}\left(\mbf \beta,
\mbf \gamma\right)&=\text{tr} \left(\bf J_{\mbf \beta}^{\left(\gamma_n\right)}\bf J_{\mbf \beta}^{-1} \bf J_{\mbf \beta}^{\left(\gamma_m\right)}\bf J_{\mbf \beta}^{-1}-\bf J_{\mbf \beta}^{\left(\gamma_n,\gamma_m\right)}\bf J_{\mbf \beta}^{-1}\right)\label{eq:bias.S.gamma}\\
\intertext{and}
\frac{\partial}{\partial \beta_m}b_{\gamma_n}\left(\mbf \beta,
\mbf \gamma\right)&=\text{tr} \left(\bf J_{\mbf \beta}^{\left(\gamma_n\right)}\bf J_{\mbf \beta}^{-1} \bf J_{\mbf \beta}^{\left(\beta_m\right)}\bf J_{\mbf \beta}^{-1}-\bf J_{\mbf \beta}^{\left(\gamma_n,\beta_m\right)}\bf J_{\mbf \beta}^{-1}\right)\label{eq:bias.S.gamma.beta}.
\end{align}
The derivatives of $\bf J_{\mbf \beta}$ are listed in Appendix 2.
\subsection{Godambe information $\bf J_\mbf \theta$}
\label{s:godambe}
For joint inference on $\mbf \theta=\left(\mbf \beta^{\top},\mbf \gamma^{\top}\right)^{\top}$ we use the asymptotic property, valid under mild regularity conditions
\begin{equation*}
\widehat{\mbf \theta} \sim N\left(\mbf \theta,\bf J_\mbf \theta^{-1}\right),
\end{equation*}
where $\bf J_\mbf \theta^{-1}=\mbf S_\mbf \theta^{-1}\mbf V_\mbf \theta \mbf S_\mbf \theta^{-\top}$, the inverse Godambe information or the sandwich estimator.

The structure of the "bread" $\mbf S_\mbf \theta^{-1}$ in the sandwich estimator is
 (\ref{eq:S.theta}), with blocks listed in (\ref{eq:S.beta.and.V.beta}), (\ref{eq:S.gamma}) and (\ref{eq:S.gamma.beta}). The lower blocks, associated with $\mbf \gamma$ are however amended with terms for bias correction as given by (\ref{eq:bias.S.gamma}) and (\ref{eq:bias.S.gamma.beta}).

The "meat" part $\mbf V_\mbf \theta$ is the variability of $\mbf \psi_\mbf \theta$ and may be structured analogously
\begin{equation}\mbf V_\mbf \theta=\begin{bmatrix} \mbf V_\mbf \beta &  \mbf V_{\mbf \beta\mbf \gamma}\\
\mbf V_{\mbf \gamma\mbf \beta}& \mbf V_\mbf \gamma
\end{bmatrix}, \label{eq:V.theta}
\end{equation}
where obviously $\mbf V_{\mbf \beta\mbf \gamma}=\mbf V_{\mbf \gamma\mbf \beta}^{\top}$.

Using (\ref{eq:S.theta}) and (\ref{eq:V.theta}) $\bf J_\mbf \theta^{-1}$ may be written as
\begin{align}
\bf J_\mbf \theta^{-1}&=
\begin{bmatrix} \bf J^\mbf \beta &  \bf J^{\mbf \beta\mbf \gamma}\\
\bf J^{\mbf \gamma\mbf \beta} & \bf J^\mbf \gamma
\end{bmatrix}\notag\\
&=\begin{bmatrix} \mbf S_\beta^{-1} &  \mbf 0\\
-\mbf S_\mbf \gamma^{-1}\mbf S_{\mbf \gamma\mbf \beta}\mbf S_\mbf \beta^{-1}& \mbf S_\mbf \gamma^{-1}
\end{bmatrix}
\begin{bmatrix} \mbf V_{\mbf \beta} &  \mbf V_{\mbf \beta\mbf \gamma}\\
\mbf V_{\mbf \gamma\mbf \beta}& \mbf V_{\mbf \gamma}
\end{bmatrix}
\begin{bmatrix} \mbf S_\mbf \beta^{-1} & -\mbf S_\mbf \gamma^{-1}\mbf S_{\mbf \gamma\mbf \beta}^{\top}\mbf S_\mbf \beta^{-1}\\
\mbf 0& \mbf S_\mbf \gamma^{-1}
\end{bmatrix}\notag\\
&=\begin{bmatrix} \mbf S_\mbf \beta^{-1}\mbf V_{\mbf \beta}\mbf S_\mbf \beta^{-1} &  \mbf S_\mbf \beta^{-1}\left(- \mbf V_{\mbf \beta}\mbf S_\mbf \beta^{-1}\mbf S_{\mbf \gamma\mbf \beta}^{\top}+\mbf V_{\mbf \gamma\mbf \beta}^{\top}\right)\mbf S_\mbf \gamma^{-1}\\
\mbf S_\mbf \gamma^{-1}\left(-\mbf S_{\mbf \gamma\mbf \beta}\mbf S_\mbf \beta^{-1}\mbf V_{\mbf \beta}+\mbf V_{\mbf \gamma\mbf \beta}\right)\mbf S_\mbf \beta^{-1}&
\mbf S_\mbf \gamma^{-1}\left(\mbf L+\mbf V_\gamma\right) \mbf S_\mbf \gamma^{-1}
\end{bmatrix}\label{eq:J.theta.inv},
\end{align}
where $\mbf L=\mbf S_{\mbf \gamma\mbf \beta}\mbf S_\mbf \beta^{-1}\left(\mbf V_{\mbf \beta}\mbf S_\mbf \beta^{-1}\mbf S_{\mbf \gamma\mbf \beta}^{\top}-\mbf V_{\mbf \gamma\mbf \beta}^{\top}\right)- \mbf V_{\mbf \gamma\mbf \beta}\mbf S_\mbf \beta^{-1}\mbf S_{\mbf \gamma\mbf \beta}^{\top}$.

The upper left block of $\bf J_\mbf \theta^{-1}$ shows that the the asymptotic variance of $\widehat{\mbf \beta}$ is unaffected by the estimation of $\mbf \gamma$. On the other hand the quantity $\mbf L$ in the lower right block represents the inflation of the asymptotic variance of $\widehat{\mbf \gamma}$ caused by the estimation of $\mbf \beta$. By (\ref{eq:S.beta.and.V.beta}) $\mbf S_\mbf \beta=-\mbf V_\mbf \beta$ and therefore the upper right block of (\ref{eq:J.theta.inv}) reduces to $\mbf S_\mbf \beta^{-1}\left(\mbf S_{\mbf \gamma\mbf \beta}^{\top}+\mbf V_{\mbf \gamma\mbf \beta}^{\top}\right) \mbf S_\mbf \gamma^{-1}$. If $\mbf S_{\mbf \gamma\mbf \beta}+\mbf V_{\mbf \gamma\mbf \beta}=\mbf 0$ then $\mbf S_\mbf \theta=-\mbf V_\mbf \theta$ and $\mbf \psi$ would have been a quasi score. In that sense the $\mbf S_{\mbf \gamma\mbf \beta}^{\top}+\mbf V_{\mbf \gamma\mbf \beta}^{\top}$ measures how much $\mbf \psi_\mbf \gamma$ deviates from being quasi-score.

Except for $\mbf V_\mbf \beta$, the blocks rely on $3$rd and $4$th moments. For practical use this seems less tractable and we will instead employ empirical variabilities of  $\breve{\mbf \psi}=\left(\mbf \psi_\mbf \beta^{\top},\breve{\mbf \psi}_\mbf \gamma^{\top}\right)^{\top}$, defined by $\widehat{\mbf V}_{\mbf \theta,\emp}=\sum_i \breve{\mbf \psi}_i(\widehat{\mbf \theta})\breve{\mbf \psi}_i(\widehat{\mbf \theta})^{\top}$.
By plugging in $\widehat{\mbf V}_{\mbf \theta,\emp}$ we obtain the empirical sandwich estimator,
$\widehat {\bf J}_{\mbf \theta,\emp}^{-1}=\mbf S_\mbf \theta^{-1}\widehat{\mbf V}_{\mbf \theta,\emp}{\mbf S}_\mbf \theta^{-1}$
\begin{align}
\widehat {\bf J}_{\mbf \theta,\emp}^{-1}&=\begin{bmatrix} \mbf S_\mbf \beta^{-1}\widehat {\mbf V}_{\mbf \beta,\emp}\mbf S_\mbf \beta^{-1} &  \mbf S_\mbf \beta^{-1}\left(-\widehat {\mbf V}_{\mbf \beta,\emp}\mbf S_\mbf \beta^{-1}\mbf S_{\mbf \gamma\mbf \beta}^{\top}+\widehat {\mbf V}_{\mbf \gamma\mbf \beta,\emp}^{\top}\right)\mbf S_\mbf \gamma^{-1}\\
\mbf S_\mbf \gamma^{-1}\left(-\mbf S_{\mbf \gamma\mbf \beta}\mbf S_\mbf \beta^{-1}\widehat {\mbf V}_{\mbf \beta,\emp}+\widehat {\mbf V}_{\mbf \gamma\mbf \beta,\emp}\right)\mbf S_\mbf \beta^{-1}&
\mbf S_\mbf \gamma^{-1}\left(\widehat {\mbf L}+\widehat {\mbf V}_{\mbf \gamma,\emp}\right) \mbf S_\mbf \gamma^{-1}
\end{bmatrix}, \label{eq:J.theta.inv.emp}
\end{align}
where $\widehat{\mbf L}=\mbf S_{\mbf \gamma\mbf \beta}\mbf S_\mbf \beta^{-1}\left(\widehat{\mbf V}_{\mbf \beta,\emp}\mbf S_\mbf \beta^{-1}\mbf S_{\mbf \gamma\mbf \beta}^{\top}-\widehat{\mbf V}_{\mbf \gamma\mbf \beta,\emp}^{\top}\right)-\widehat{\mbf V}_{\mbf \gamma\mbf \beta,\emp}\mbf S_\mbf \beta^{-1}\mbf S_{\mbf \gamma\mbf \beta}^{\top}$.

We may replace $\widehat {\mbf V}_{\mbf \beta,\emp}$ by $\mbf V_{\mbf \beta}=\mbf S_\mbf \beta^{-1}$ to obtain the semi-empirical sandwich estimator
\begin{align}
{\widehat {\bf J}}_{\mbf \theta,\textsc{\scriptsize Sem}}^{-1}&=
\begin{bmatrix} -\mbf S_\mbf \beta^{-1} &  \mbf S_\mbf \beta^{-1}\left(\mbf S_{\mbf \gamma\mbf \beta}^{\top}+\widehat { \mbf V}_{\mbf \gamma\mbf \beta,\emp}^{\top}\right)\mbf S_\mbf \gamma^{-1}\\
\mbf S_\mbf \gamma^{-1}\left(\mbf S_{\mbf \gamma\mbf \beta}+\widehat {\mbf V}_{\mbf \gamma\mbf \beta,\emp}\right)\mbf S_\mbf \beta^{-1}&
\mbf S_\mbf \gamma^{-1}\left(\widetilde{\mbf L}+\widehat{\mbf  V}_{\mbf \gamma,\emp}\right) \mbf S_\mbf \gamma^{-1}
\end{bmatrix}\notag,
\end{align}
where now $\widetilde {\mbf L}=-\mbf S_{\mbf \gamma\mbf \beta}\mbf S_\mbf \beta^{-1}\left(\mbf S_{\mbf \gamma\mbf \beta}^{\top}+\widehat
{\mbf V}_{\mbf \gamma\mbf \beta,\emp}^{\top}\right)-\widehat {\mbf V}_{\mbf \gamma\mbf \beta,\emp}\mbf S_\mbf \beta^{-1}\mbf S_{\mbf \gamma\mbf \beta}^{\top}$.

The bias correction  $\mbf b_{\mbf \gamma}\left(\mbf \gamma,\mbf \beta\right)$ applied to $\mbf \psi_\mbf \gamma$ causes that part of $\breve{\mbf \psi}$ to have a non-zero mean value.
\section{Simulation}
\label{s:simulation}
Some key properties of our method were addressed by an extensive simulation study in which 4000 data sets were simulated for each of 36 different configurations specified by combinations of the following Tweedie parameters: $r_1 \in \left\{1\cd5, 2\cd0, 2\cd5, 3\cd0\right\}, r_2 \in \left\{1\cd5, 2\cd0, 3\cd0\right\}$ and $r_3 \in \left\{1\cd0, 1\cd5, 2\cd0, 3\cd0\right\}$. Table \ref{tbl:simconfig} lists the model parameter settings used for the simulations. These settings are repeated across all combination of the latent variables Tweedie parameters $r_1$ and $r_2$. The data sets were simulated with $15$ clusters each, and each cluster consisting of latent \AR{1} time series of length $30$ and a log-linear regression for the fixed part. The correlation parameter $\alpha$ and the response dispersion parameter $\rho^2$ varied with the response model. Except for the Poisson case, for which we used an intercept of $4\cd0$, all other intercepts and all slopes were identical across scenarios. For other response models than the Poisson, the parameters were chosen to attain similar second moments for the marginal response. The coefficient of variation of the marginal response, based on these settings, ranged from $0\cd497$ (Poisson) to $0\cd579$ (Gamma).

As $\psi_\beta$ is a quasi score estimating function, with well known and optimal properties the simulation study naturally focuses on the estimates of the association parameters.
\begin{table}[hptb]
\noindent
\caption{Parameter settings used for the simulations.}\label{tbl:simconfig}
\begin{center}
\begin{tabular}{l c c c c c c c}
\toprule
\multicolumn{2}{l}{Response Model}&\multicolumn{2}{c}{Regression}&\multicolumn{3}{c}{ Dispersion}&\multicolumn{1}{c}{Correlation}\\
\cmidrule(lr{.6em}){3-4}\cmidrule(lr{.6em}){5-7}\cmidrule(lr{.6em}){8-8}
\multicolumn{1}{l}{Distribution}&\multicolumn{1}{c}{$r_3$}&\multicolumn{1}{c}{$\beta_0$}&\multicolumn{1}{c}{$\beta_1$}
&\multicolumn{1}{c}{$\sigma^2$}&\multicolumn{1}{c}{$\omega^2$}&\multicolumn{1}{c}{$\rho^2$}&\multicolumn{1}{c}{$\alpha$}\\
\hline
Poisson&1$\cdot$0&4$\cdot$0&0$\cdot$3&0$\cdot$05&0$\cdot$15&1$\cdot$0000&0$\cdot$40\\
Compound Poisson&1$\cdot$5&1$\cdot$6&0$\cdot$3&0$\cdot$05&0$\cdot$15&0$\cdot$1175&0$\cdot$55\\
Gamma&2$\cdot$0&1$\cdot$6&0$\cdot$3&0$\cdot$05&0$\cdot$15&0$\cdot$0850&0$\cdot$50\\
Inverse Gaussian&3$\cdot$0&1$\cdot$6&0$\cdot$3&0$\cdot$05&0$\cdot$15&0$\cdot$0200&0$\cdot$40\\
\bottomrule
\end{tabular}
\end{center}
\end{table}
\subsection*{Robustness}
Simulations with varying configurations of $r_1$ and $r_2$ was used for studying the assumed robustness against the lack of knowledge about the Tweedie parameters driving the latent process.
Along the same lines we investigated how the model performed across an appropriate range of the Tweedie parameter $r_3$ for the response variable.

\begin{figure}
\centering
\includegraphics[scale=0.37]{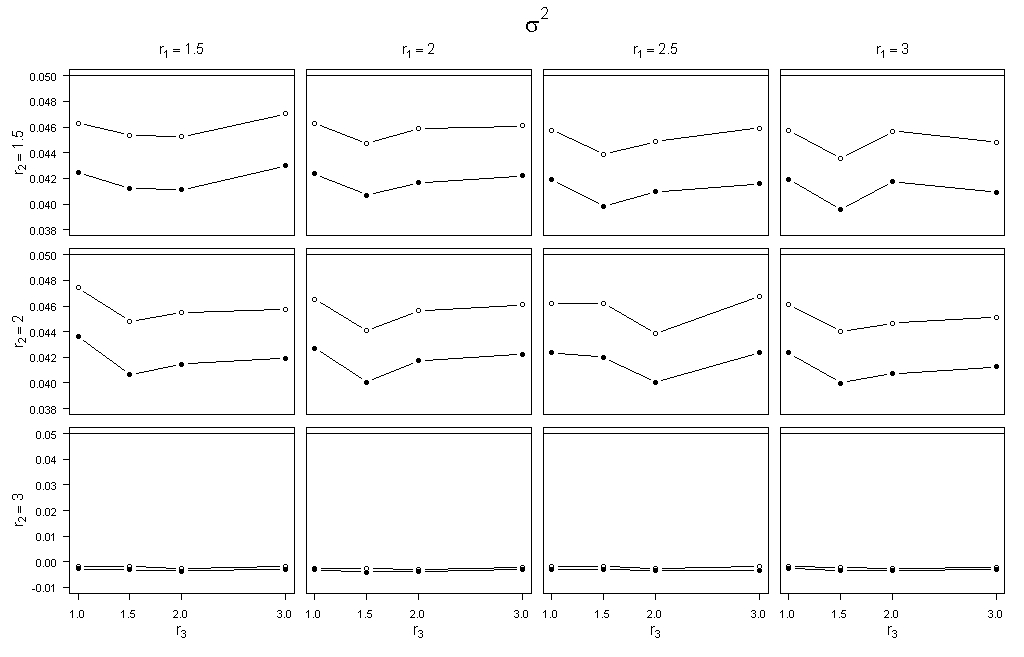}
\caption{Median values of $\widehat{\sigma}^2$ for bias-corrected ($\circ$) and un-corrected  ($\bullet$) estimation compared with the true value $\sigma^2=0\cd05$ (horizontal line).}
\label{fig:sigma2sim}
\end{figure}

Figure \ref{fig:sigma2sim} shows the median values of estimates of $\sigma^2$, for all combinations of the Tweedie parameter considered. Corresponding plots for $\omega^2$, $\rho^2$ and $\alpha$,  not included in the article, show similar patterns across the range of $r_1$ values considered. Also there seems very little difference in the patterns across the range of $r_2$ values for $\widehat\rho^2$ and $\widehat\alpha$, whereas $\widehat\sigma^2$ and $\widehat\omega^2$ show a markedly higher bias for $r_2=3$ than for $r_2=1\cd5, 2$. Our approach hence appear reasonably robust against varying specifications of the latent process.
The asymptotic variances of the estimates of the association parameters (\ref{eq:J.theta.inv.emp}), enables us to compute estimates of coverage probabilities for $95\%$ aymptotic confidence intervals, based on these.
These coverage probabilities are listed in Table \ref{tbl:summary.covr}.
\begin{table}[hptb]
\caption{Summary of coverages for 95 \% confidence intervals across varying configurations.}\label{tbl:summary.covr}
\begin{center}
\begin{tabular}{l c c c c c c}
\toprule
\multicolumn{1}{l}{Statistic}&\multicolumn{1}{c}{$\beta_0$}&\multicolumn{1}{c}{$\beta_1$}&\multicolumn{1}{c}{$\sigma^2$}&\multicolumn{1}{c}{$\omega^2$}&\multicolumn{1}{c}{$\rho^2$}&\multicolumn{1}{c}{$\alpha$}\\ \hline
1st~Qu.&$73\cd54$\%~~&$91\cd62$\%~~&$99\cd95$\%~~&$97\cd55$\%~~&$99\cd12$\%~~&$98\cd91$\%~~\\
Median~&$92\cd78$\%~~&$94\cd29$\%~~&$100\cd00$\%~~&$98\cd98$\%~~&$99\cd44$\%~~&$99\cd32$\%~~\\
3rd~Qu.&$93\cd09$\%~~&$94\cd68$\%~~&$100\cd00$\%~~&$99\cd83$\%~~&$99\cd66$\%~~&$99\cd85$\%~~\\
\bottomrule
\end{tabular}
\end{center}
\end{table}
While the coverages indicate the standard errors of the regression parameters to be precise, the coverages are consistently much too big for the association parameters, indicating overestimation of their standard errors.
\subsection*{Bias correction}
The magnitude of the nuisance parameter bias correction is assessed by a duplicate analysis of the simulated data sets, except that the second estimation is done without bias correction.

Within the range of configurations considered here, there were generally little bias on the average of the estimated association parameters. Apart from a few minor exceptions, that may well be referred to sampling error, the bias correction pulled the estimates closer to their true values. The bottom level dispersion parameter $\sigma^2$ shows markedly the highest correction.

\section{Data Analysis}
\label{s:data.analysis}
Knowledge about the growth of fish is important for the assessment of fish biomass. For this purpose, many fisheries management programmes sample otoliths on a regular basis. An otolith is a structure located in the inner ear of fish and is built by deposit of calcium carbonate, protein and a variety of trace elements. It carries information about age and growth patterns, by means of alternating opaque and translucent bands.
When viewed in transmitted light, a translucent band represents a low level of deposition of proteins in the calcium carbonate crystal structure corresponding to a period of slow growth \citep{Mosegaard1987}.
Sub-seasonal bands, representing daily cycles, can sometimes be identified within the annual bands \citep{Pannella1971}.

\begin{figure}
\centering
\includegraphics[scale=0.37]{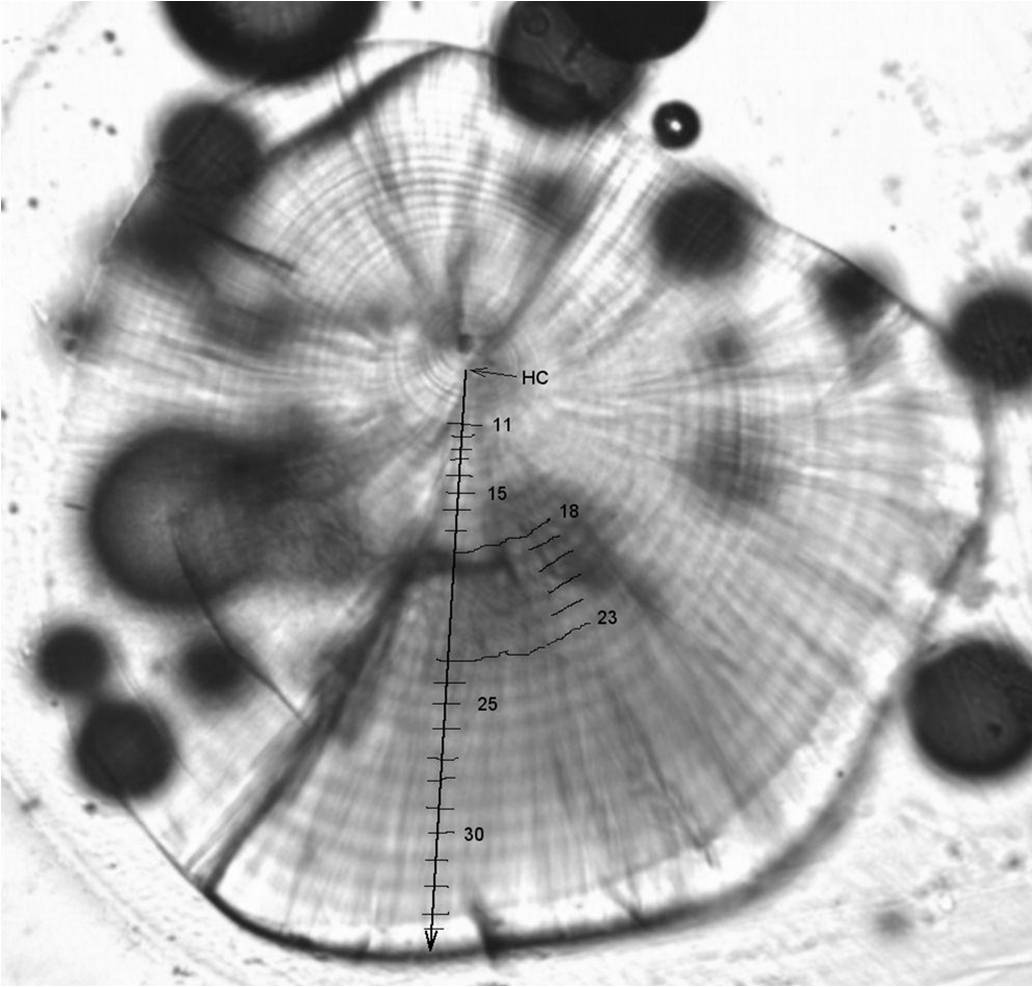}
\caption{A photo of an otolith with radius and band marks. The trace marks for band 18--23 are identified outside the radius and only the span of these bands can be measured. HC: Hatch Check.}
\label{fig:OtolithFoto}
\end{figure}

The data collected and analysed by \cite{Clausen2007} contains measurements of daily growth bands of otoliths collected from juvenile herrings ({\it Clupea harengus}). The age in days was determined, by counting bands, for each of the sampled specimen and along with the time of sampling they were categorized as being offspring from one of three spawner types: autumn, winter and spring. These are distinct stock components but mix on the nursery and feeding grounds. For stock assessment purposes, it is of interest to be able to discriminate between them. \cite{Clausen2007} used otolith characteristics for this purpose.

With each fish being a cluster and the sequence of bands within fish giving the longitudinal structure, otoliths measurements are amenable to our framework.
We analysed the data from \cite{Clausen2007} to illustrate the use of our model for this type of data.

For compatibility across the collection of otoliths, the band widths are measured along similar radii on all otoliths. If the band marks are not all clearly identifiable along this transect, two or more adjacent bands are aggregated and the total width of these bands is taken (Fig. \ref{fig:OtolithFoto}). The count of bands between two measurement marks is then based on intermediate band marks identified elsewhere on the otoliths. It is common practice to use the average of such aggregated bands, in place of correct measurements; a feature also found in the present data. To avoid successive values obtained from the same aggregation of bands, it sufficed to sub-sample every 8th value for our analysis.

The sampled fish have different ages and therefore display differences in the lengths of their band width series. To avoid bias caused by data selection, we truncated the sequences of band measurements to the shortest sequence within each spawning category. Furthermore the first 10 bands were left out of the analysis, as their measurements were considered too imprecise.

\begin{figure}
\centering
\includegraphics[scale=0.45]{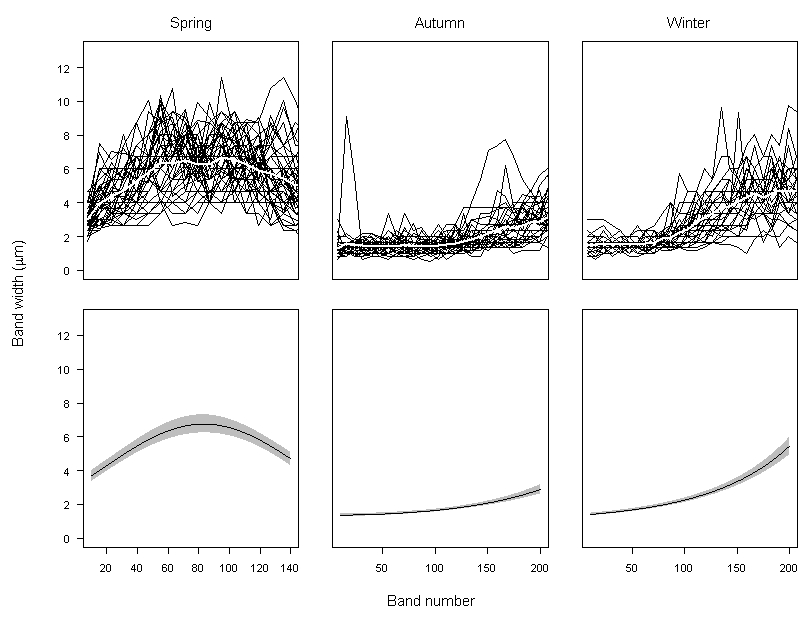}
\caption{Width measurements at every 8th band. Thick white line indicates average measurements over otoliths. Top row: observed, bottom row: estimated.}
\label{fig:observedandfittedwidth}
\end{figure}

Two variant models were estimated: one fixing the response Tweedie parameter $r_3=2$, corresponding to a gamma distribution and the other one estimating this parameter. Judging from exploratory plots (Fig. \ref{fig:observedandfittedwidth}) of the observed width measurements the fixed part could appropriately be modeled as 2nd order polynomials of the bands and with potential different coefficients for the three seasons: \texttt{width $\sim$ (band+band$^2$)*season} using the log link.

The initial models contained 9 regression parameters. Autumn was chosen as base level for the season factor, to enable a direct comparison between autumn and winter, as these appeared most alike among the the three seasons. The models were reduced to final models, with 6 regression parameters, through a succession of Walds test and re-estimations. Based on inspection of the auto-correlation and the partial auto-correlation function for individual otoliths, an \AR{1} model was deemed appropriate. Estimates for the parameters of the final models and standard errors are listed in Table \ref{tbl:parmest}.

\begin{table}[hptb]
\caption{Parameter estimates, standard errors (SE) and $p$-values for both fixed and estimated $r_3$ models. * The $p$-value for $r_3$ applies to the hypothesis H$_0:\, r_3=2$.
{\bf aut}: autumn; {\bf win}: winter; {\bf spr}: spring.
}
\label{tbl:parmest}
\begin{center}
\begin{tabular}{l c c c c c c c} 
\toprule
\multicolumn{1}{l}{}&\multicolumn{3}{c}
{Fixed $r_3=2$}&\multicolumn{1}{l}{\,}&\multicolumn{3}{c}{Estimated $r_3$}\\
\cmidrule(lr{.6em}){2-4}\cmidrule(lr{.6em}){6-8}
\multicolumn{1}{l}{Parameter}&\multicolumn{1}{c}{Est}&\multicolumn{1}{c}{SE}&\multicolumn{1}{c}
{$p$-value}&\multicolumn{1}{l}{}&\multicolumn{1}{c}{Est}&\multicolumn{1}{c}{SE}&\multicolumn{1}{c}{$p$-value}\\
\hline
$\beta_\texttt{aut+win}$&$0\cd3118$&$0\cd0292$&$<0\cd0001$&&$0\cd3113$&$0\cd0292$&$<0\cd0001$\\
$\beta_\texttt{spr}$&$1\cd1304$&$0\cd0553$&$<0\cd0001$&&$1\cd1288$&$0\cd0556$&$<0\cd0001$\\
$\beta_\texttt{spr:band}$&$0\cd0187$&$0\cd0016$&$<0\cd0001$&&$0\cd0187$&$0\cd0016$&$<0\cd0001$\\
$\beta_\texttt{win:band}$&$0\cd0032$&$0\cd0003$&$<0\cd0001$&&$0\cd0032$&$0\cd0003$&$<0\cd0001$\\
$\beta_{\texttt{(aut+win):band}^2}$&$1\cd9 \times 10^{-5}$&$1\cd4 \times 10^{-6}$&$<0\cd0001$&&$1\cd9 \times 10^{-5}$&$1\cd4 \times 10^{-6}$&$<0\cd0001$\\
$\beta_{\texttt{spr:band}^2}$&$\hspace{-6pt}-0\cd0001$&$9\cd9 \times 10^{-6}$&$\hspace{10pt}0\cd 0001$&&$\hspace{-6pt}-0\cd0001$&$1\cd0 \times 10^{-5}$&$<0\cd0001$\\
$\sigma^2$&$0\cd0040$&$0\cd1639$&$\hspace{10pt}0\cd9807$&&$0\cd0042$&$0\cd1627$&$\hspace{10pt}0\cd9796$\\
$\omega^2$&$0\cd0277$&$2\cd6770$&$\hspace{10pt}0\cd9918$&&$0\cd0275$&$3\cd4696$&$\hspace{10pt}0\cd9937$\\
$\rho^2$&$0\cd0153$&$4\cd1817$&$\hspace{10pt}0\cd9971$&&$0\cd0115$&$7\cd0517$&$\hspace{10pt}0\cd9987$\\
$\alpha$&$0\cd8321$&$0\cd0823$&$<0\cd0001$&&$0\cd8325$&$6\cd7673$&$\hspace{10pt}0\cd9021$\\
$r_3$&&&&&$2\cd2700$&$1\cd7627$&$\hspace{16pt}0\cd8783^*$\\
\bottomrule
\end{tabular}
\end{center}
\end{table}

The two regressions are estimated almost exactly the same, whether $r_3$ is estimated or assumed known. This reflects the $\gamma$ insensitivity of $\mbf \psi_\mbf \beta$.
From the fit we conclude that autumn and winter differ only by the 1st order term whereas autumn and spring differ by all three terms. The fitted curves from the fixed-$r_3$ model are plotted in Fig. \ref{fig:observedandfittedwidth}.

The association parameter estimates were very similar for the two models, but giving quite different standard errors. The rather large standard errors for the dispersion parameter estimates confirms the impression from the simulations in Section \ref{s:simulation}, that the use of empirical variabilities in the sandwich estimator (\ref{eq:J.theta.inv.emp}) may lead to standard errors too big to be of any practical use. This seems to be the cost of avoiding the use of higher moments . The standard errors for the correlation parameter $\alpha$ in the fixed-$r_3$ case appears to be more realistic but raises dramatically when $r_3$ is estimated. In that case the standard error for $r_3$ seems moderate. The $\alpha$ parameter applies on the scale of the sub-sampling frequency. A back calculation to a day-to-day serial correlation and assuming the \AR{1} model, leads to a value of about $0\cd97.$
\section{Discussion}
\label{s:discussion}
There are a few useful extensions of the method worth mentioning here. It
would be straightforward to extend the model to multiple levels of random
effects, adding further levels by repeated use of conditional Tweedie
distributions. A second useful extension would be to allow regression
modelling for the association parameters, along the lines of \cite{Davidian1987}.

Finally, the model can be easily adapted to handle multivariate data. By use of
conditionally independent Poisson response variables, the model can be extended to
binomial or categorical data, leading
to beta-binomial-like or Dirichlet-multinomial-like models. This topic is addressed in a companion article
currently in preparation.

\section*{Acknowledgements}
The authors thank Rob Fryer for valuable comments on our approach and
Lotte Wors\o e Clausen and Helge Paulsen for providing us with the otolith data and guidance in understanding them.

\appendix
\section*{Appendix 1}
\subsection*{Covariance structure}
\label{app:covarstruct}
From the model specification (\ref{eq:Zi})--(\ref{eq:Yitj}) we derive the the marginal covariance between two observations within the $i$th cluster. This is done in three steps by means of the law total of variance. From $\mE Z_i = 1$ and $\mVar{Z_i}=\sigma^2$ we first get
\begin{align*}
\mCov{Z_{it}}{Z_{it'}}&=
E\left\{\mCov{Z_{it}}{Z_{it'}\mid Z_*}\right\}+ \mCov{\mE{Z_{it}\mid Z_*}}{\mE{Z_{it'}\mid Z_*}}\\
&=\delta_{t}^{t'}\mE{\omega^2 Z_i} + \aplus{-2}\mVar{Z_i}\\
&=\delta_{t}^{t'}\omega^2
+ \aplus{-2}\sigma^2,
\end{align*}
from which we obtain
\begin{align*}
\mCov{Q_{it}}{Q_{it'}}
&=\sum_{s=0}^\infty \sum_{s'=0}^\infty \alpha_s \alpha_{s'} \,\mCov{Z_{i\,t-s}}{Z_{i\,t'-s'}}\\
&=\omega^2\sum_{s=0}^\infty \alpha_s \alpha_{s+|t-t'|} +\sigma^2,
\end{align*}
and finally arrive at
\begin{align*}
\mCov{Y_{it}}{Y_{i't'} }&=
E\left\{\mCov{Y_{it}}{Y_{it'}\mid Z}
\right\}+ \text{cov}\left\{\mE{Y_{it}\mid
Z},\mE{Y_{i't'}\mid
Z}\right\}\\
&=\delta_{t}^{t'}\mVar{Y_{it}} + \mu_{it}\mu_{ij'}\,\mCov{Q_{it}}{Q_{it'}}\\
&=\delta_{t}^{t'}\,\mu_{it}^{r_{3}}\,\rho^2+\mu_{it}\mu_{it'}\,\left(\omega^2
\sum_{s=0}^\infty \alpha_s \alpha_{s+|t-t'|} +
\sigma^2
\right)
\end{align*}

\section*{Appendix 2}
\subsection*{Process Correlation Matrix $K\left(\alpha\right)$ for \MA{q} and \AR{1} processes}
The latent process linear filter $Q_{it}=\sum_{s=0}^\infty \alpha_s Z_{i\,t-s}$,
induces the process correlation matrix $K\left(\alpha\right)$, with $tt'$th entry
$\left\{ K\left(\alpha\right)\right\}_{tt'}=\sum_{s=0}^\infty \alpha_s \alpha_{s+|t-t'|}.$

An \MA{q} process is given by $\alpha_0=1$ and $\alpha_s=0$ for $s>q$ and has first and second derivative matrices with $tt'$th entries given by
\begin{align*}
\left\{\frac{\partial}{\partial \alpha_k} K\left(\alpha\right)\right\}_{tt'}&=
\alpha_{k+|t-t'|}\delta_{k\leq q-|t-t'|}+\alpha_{k-|t-t'|}\delta_{k\geq |t-t'|}\\
\intertext{and}
\left\{\frac{\partial^2}{\partial \alpha_k\partial \alpha_m} K( \alpha)\right\}_{tt'}&=
\delta_t^{t'}\delta_k^{m}+\delta_{|t-t'|}^{|k-m|}
\end{align*}
respectively. Here $\delta_t^{t'}$, $\delta_{k\leq q-|t-t'|}$ etc. are variant forms of the Kronecker delta with obvious definitions.

An \AR{1} process is given by $\alpha_s=\alpha^s$; $\alpha \in (0,1)$, from which we get the $tt'$th entry of $ K\left(\alpha\right)$
\begin{equation*}
\left\{ K\left(\alpha\right)\right\}_{tt'}=\sum_{s=0}^\infty \alpha^{2s+|t-t'|}=\alpha^{|t-t'|}\sum_{s=0}^\infty \alpha^{2s}=\frac{\alpha^{|t-t'|}}{1-\alpha^2}.
\end{equation*}
Consequently the $k$th sub- and super-diagonal of $\frac{\partial}{\partial\alpha}{K\left(\alpha\right)}$ and $\frac{\partial^2}{\partial\alpha^2}{ K\left(\alpha\right)}$ have elements
\begin{align*}
\frac{\partial}{\partial\alpha}\left(\frac{\alpha^k}{1-\alpha^2}\right)
&=\frac{\left(2-k\right)\alpha^{k+1}+k\alpha^{k-1}}{\left(1-\alpha^2\right)^2}\\
\intertext{and}
\frac{\partial^2}{\partial\alpha^2}\left(\frac{\alpha^k}{1-\alpha^2}\right)
&=\frac{\left(k^2-5k+6\right)\alpha^{\left(k+2\right)}+\left(-2k^2+6k+2\right)\alpha^k+\left(k-1\right)k\alpha^{\left(k-2\right)}}{\left(1-\alpha^2\right)^3}
\end{align*}
respectively.

\section{Appendix 3}
\subsection*{Derivatives of $J_{\beta}$}
\label{app:derivatives.godambe.beta}
The derivatives of $J_{\beta}$ involved in calculating the $\gamma$- and $\beta$- sensitivities of $\breve{ \psi}_{ \gamma}\left(\beta,\gamma\right)$ are
\begin{align*}
J_{\beta}^{\left(\gamma_n\right)}&=-\sum_{i=1}^I  D_i^\top W_{in}  D_i\\
J_{\beta}^{\left(\beta_m\right)}&=\sum_{i=1}^I  D_i^{\left(\beta_m\right)\top}C_{i}^{-1} D_i+
 D_i^\top C_{i}^{-1} D_i^{\left(\beta_m\right)}-
 D_i^\top C_{i}^{-1} C_{i}^{\left(\beta_m\right)} C_{i}^{-1} D_i\\
J_{\beta}^{\left(\gamma_n,\gamma_m\right)}&=\sum_{i=1}^I  D_i^{\top}\left(W_{im}C_iW_{in}+
W_{in}C_iW_{im}-C_{i}^{-1}C_{i}^{\left(\gamma_n,\gamma_m\right)}C_{i}^{-1}\right)D_i\\
J_{\beta}^{\left(\gamma_n,\beta_m\right)}&=-\sum_{i=1}^I D_i^{\left(\beta_m\right)\top}W_{in}D_i+
D_i^{\top}W_{in}D_i^{\left(\beta_m\right)}+D_i^{\top}W_{in}^{\left(\beta_m\right)}D_i,
\end{align*}
where $ W_{in}=-\frac{\partial}{\partial \gamma_n} C_i^{-1}= C_i^{-1}\left(\frac{\partial}{\partial \gamma_n} C_i\right) C_i^{-1}$ and $D_i^{\left(\beta_m\right)}=\frac{\partial}{\partial \beta_m}D_i$ etc. 

\bibliographystyle{biom}
\bibliography{rene}

\end{document}